# Revisiting the Coulomb problem: A novel representation of the confluent hypergeometric function as an infinite sum of discrete Bessel functions

A. D. Alhaidari

*Saudi Center for Theoretical Physics, P.O. Box 32741, Jeddah 21438, Saudi Arabia*

**Abstract:** We use the tridiagonal representation approach to solve the radial Schrödinger equation for the continuum scattering states of the Coulomb problem in a complete basis set of discrete Bessel functions. Consequently, we obtain a new representation of the confluent hypergeometric function as an infinite sum of Bessel functions, which is numerically very stable and more rapidly convergent than another well-known formula.

**Keywords**: confluent hypergeometric function, Bessel function, Coulomb problem, tridiagonal representations, recursion relation

## 1. Introduction

The Bessel function, $J_\nu(z)$, for a general parameter and argument is one of the most studied special functions in the mathematics and physics literature. It appears in the solution of many problems in science and engineering and so does the confluent hypergeometric function $_1F_1(a;b;z)$. Connecting these two special functions by some mathematical relations that are numerically stable and convergent has always been of great interest to researchers in various fields. Here, we use the tridiagonal representation approach (TRA) [1] to solve the Coulomb scattering problem in a complete set of Bessel functions. Comparing our solution to the well-known Coulomb wavefunction, we derive a new representation of $_1F_1(a;b;z)$ as an infinite sum of discretized Bessel functions. It is numerically very stable and more rapidly convergent than another well-known formula, especially asymptotically. For details on how to use the TRA in solving quantum mechanical problems, one may consult [1-3] and references therein. However, no such particulars are prerequisites for understanding the current work.

## 2. TRA solution of the Coulomb problem

We want to use the TRA to solve the following radial Schrödinger equation for Coulomb scattering at some positive energy $E$ and angular momentum $\ell$ in a special basis

$$\left[ -\frac{1}{2}\frac{d^2}{dr^2} + \frac{\ell(\ell+1)}{2r^2} + \frac{Z}{r} \right]\psi(r) = E\psi(r), \qquad (1)$$

where $Z$ is the electric charge and we have adopted the atomic units $\hbar = m = 1$. In the TRA, we start by expanding the solution in a series using a complete set of functions $\{\phi_n\}_{n=0}^\infty$. That is, we write $\psi(r) = \sum_{n=0}^\infty f_n \phi_n(\lambda r)$, where $\{f_n\}$ are expansion coefficients that depend on the physical parameters and energy, and $\lambda$ is a positive scale parameter of inverse length dimension. Once a basis set $\{\phi_n\}_{n=0}^\infty$ is chosen, all physical information about the system is contained in the set $\{f_n\}$ [1-3]. Rigorously, one has to show that the series is bounded and convergent in the relevant domain. Nonetheless, for the present work such properties of the resulting series will be self-evident and numerically obvious. We choose $\phi_n(x) = x^\mu J_{n+\nu}(x)$, where $x = \lambda r$, $\mu$ and $\nu$ are dimensionless parameters to be determined but such that $\nu > 0$. Writing Eq. (1) as $\mathcal{D}\psi(r) = 0$, where $\mathcal{D}$ is the wave operator, turns it into the sum $\sum_n f_n \mathcal{D}\phi_n(x) = 0$. The action of the wave operator on $\phi_n(x)$ in terms of the new dimensionless variable $x$ reads as follows

$$\mathcal{D}\phi_n(x) = -\frac{\lambda^2}{2} x^{\mu-1}$$
$$\left\{ (2\mu-1)\frac{d}{dx} - x + \frac{1}{x}\left[ (n+\nu)^2 + \left(\mu-\tfrac{1}{2}\right)^2 - \left(\ell+\tfrac{1}{2}\right)^2 \right] - 2\frac{Z}{\lambda} + \frac{k^2}{\lambda^2} x \right\} J_{n+\nu}(x) \qquad (2)$$

where $k = \sqrt{2E}$ and we have used the differential equation for the Bessel function that reads [4]

$$\left[ x^2 \frac{d^2}{dx^2} + x\frac{d}{dx} + x^2 - (n+\nu)^2 \right] J_{n+\nu}(x) = 0. \qquad (3)$$

In the TRA, it is required that $\mathcal{D}\phi_n(x)$ be of the following form [1-3]

$$\mathcal{D}\phi_n(x) = \omega(x)\left[ \alpha_n \phi_n(x) + \beta_{n-1} \phi_{n-1}(x) + \gamma_n \phi_{n+1}(x) \right], \qquad (4)$$

where $\{\alpha_n, \beta_n, \gamma_n\}$ are $x$-independent parameters and $\omega(x)$ is an entire function, which is nodeless inside the $x$ interval. Equation (4) is referred to as the "fundamental TRA constraint". Consequently, the wave equation (1), which reads $\sum_n f_n \mathcal{D}\phi_n(x) = 0$, becomes a three-term recursion relation for the expansion coefficients of the wavefunction that reads

$$\alpha_n f_n + \beta_n f_{n+1} + \gamma_{n-1} f_{n-1} = 0, \qquad (5)$$

for $n = 0, 1, 2, \ldots$. It is solvable for all $\{f_n\}$ starting with the two initial values $f_{-1} := 0$ and $f_0$. Hence, a solution of the differential wave equation (1) is equivalent to the solution of the discrete algebraic equation (5), which is the main feature of the TRA as an algebraic method. To apply the fundamental TRA constraint (4) to Eq. (2), we need to employ the following differential property and recursion relation of the Bessel function [4]

$$\frac{d}{dx} J_{n+\nu}(x) = \frac{1}{2}\left[ J_{n-1+\nu}(x) - J_{n+1+\nu}(x) \right], \qquad (6)$$

$$\frac{1}{x} J_{n+\nu}(x) = \frac{1/2}{n+\nu}\left[ J_{n+1+\nu}(x) + J_{n-1+\nu}(x) \right]. \qquad (7)$$

Now, the fundamental TRA constraint (4) and these two relations require that the two terms proportional to $x$ inside the curly brackets of Eq. (2) must be eliminated. Therefore, we must choose the basis scale parameter $\lambda$ as $\lambda^2 = k^2$, which turns Eq. (2) into the following

$$\mathcal{D}\phi_n(x) = -E x^{\mu-1}\left\{ (2\mu-1)\frac{d}{dx} + \frac{1}{x}\left[ (n+\nu)^2 + \left(\mu-\tfrac{1}{2}\right)^2 - \left(\ell+\tfrac{1}{2}\right)^2 \right] - 2\sigma \right\} J_{n+\nu}(x). \qquad (8)$$

where $\sigma = Z/k$. To simplify even further, we take the basis parameter $\mu = \frac{1}{2}$ turning this equation, after the use of the recursion (7), into the following

$$\mathcal{D}\phi_n(x) = \frac{-E}{2x}\left\{-4\sigma\phi_n(x) + \frac{1}{n+\nu}\left[(n+\nu)^2 - \left(\ell+\tfrac{1}{2}\right)^2\right][\phi_{n+1}(x) + \phi_{n-1}(x)]\right\}, \quad (9)$$

where $\phi_n(x) = x^\mu J_{n+\nu}(x) = \sqrt{kr}\, J_{n+\nu}(kr)$. Now, this equation is identical to the fundamental TRA constraint (4) with $\omega(x) = -E/2x$ and

$$\alpha_n = -4\sigma, \qquad \gamma_n = \beta_{n-1} = (n+\nu) - \frac{\left(\ell+\tfrac{1}{2}\right)^2}{n+\nu}. \quad (10)$$

Therefore, the algebraic equation (5), which is equivalent to the original differential wave equation (1), becomes the following three-term recursion relation for the expansion coefficients of the wavefunction

$$y P_n(y) = \left[(n+\nu+1) - \frac{\left(\ell+\tfrac{1}{2}\right)^2}{n+\nu+1}\right]P_{n+1}(y) + \left[(n+\nu-1) - \frac{\left(\ell+\tfrac{1}{2}\right)^2}{n+\nu-1}\right]P_{n-1}(y), \quad (11)$$

where $y := 4\sigma$ and we wrote $f_n(E) := f_0(E)P_n(y)$ making $P_0(y) = 1$. This recursion relation is solvable for all $\{P_n(y)\}_{n=0}^\infty$ as polynomials in $y$ starting with the two initial values $P_{-1}(y) := 0$ and $P_0(y) = 1$. Finally, the Coulomb wavefunction becomes

$$\psi(r) = f_0(E)\sqrt{kr}\sum_{n=0}^\infty P_n(4\sigma)J_{n+\nu}(kr). \quad (12)$$

The energy factor $f_0(E)$ is determined from the boundary conditions. Nonetheless, using our findings in the next section, we obtain $\nu = \ell + \tfrac{1}{2}$ and

$$f_0(E) = \frac{\sqrt{\pi/2}}{\Gamma(\ell+1)}e^{-\pi\sigma/2}\left|\Gamma(\ell+1\pm i\sigma)\right|. \quad (13)$$

## 3. Representation of the confluent hypergeometric function as an infinite sum of discrete Bessel functions

The Coulomb wavefunction that solves Eq. (1) is well known and is written as (see, for example, Ref. [5] and Ref. [6])

$$\psi(r) = \frac{2^\ell e^{-\pi\sigma/2}}{\Gamma(2\ell+2)}\left|\Gamma(\ell+1\pm i\sigma)\right|(kr)^{\ell+1}e^{\pm ikr}\,{}_1F_1(\ell+1\pm i\sigma;2\ell+2;\mp 2ikr), \quad (14)$$

By equating this to our TRA solution (12), we obtain directly the following representation of the confluent hypergeometric function as an infinite sum of Bessel functions

$${}_1F_1(a;b;z) = A e^{z/2}(iz/2)^{(1-b)/2}\sum_{n=0}^\infty P_n(2i(b-2a))J_{n+\nu}(iz/2). \quad (15)$$

For the Coulomb problem, $a = \ell+1+i\sigma$, $b = 2\ell+2$, and $z = -2ikr$. The constants $A$ and $\nu$ are functions of $a$ and/or $b$ to be determined by evaluating (15) at some special values. For example, taking $b = 2a$ and using the well-known identity (see, for example, page 283 in [4])

$$_1F_1(a;2a;z) = \Gamma(a+\tfrac{1}{2})(z/4i)^{\frac{1}{2}-a} e^{z/2} J_{a-\frac{1}{2}}(z/2i), \qquad (16)$$

we deduce

$$\Gamma(a+\tfrac{1}{2})2^{a-\tfrac{1}{2}} J_{a-\frac{1}{2}}(z/2i) = A \sum_{n=0}^{\infty} P_n(0) J_{n+\nu}(z/2i), \qquad (17)$$

which implies that $P_n(0) = \delta_{n,0}$. This result together with the recursion relation (11) dictate that $\nu = \ell + \tfrac{1}{2} = \tfrac{b-1}{2}$ and makes $A = 2^\nu \Gamma(\nu+1)$. Therefore, we obtain finally the following representation

$$_1F_1(a;b;z) = 2^{(b-1)/2} \Gamma\left(\tfrac{b+1}{2}\right) e^{z/2} (iz/2)^{(1-b)/2} \sum_{n=0}^{\infty} P_n(2i(b-2a)) J_{n+\frac{b-1}{2}}(iz/2), \qquad (18a)$$

where

$$y P_n(y) = \frac{1}{2}\left[(2n+b+1) - \frac{(b-1)^2}{2n+b+1}\right] P_{n+1}(y) + \frac{1}{2}\left[(2n+b-3) - \frac{(b-1)^2}{2n+b-3}\right] P_{n-1}(y), \qquad (18b)$$

for $n = 0,1,2,...$ with $y = 2i(b-2a)$, $P_0 = 1$ and $P_{-1} := 0$. Figure 1 illustrates the convergence of the series (18a) for an extended range of the argument $x$. On the other hand, its accuracy is demonstrated in Table I below.

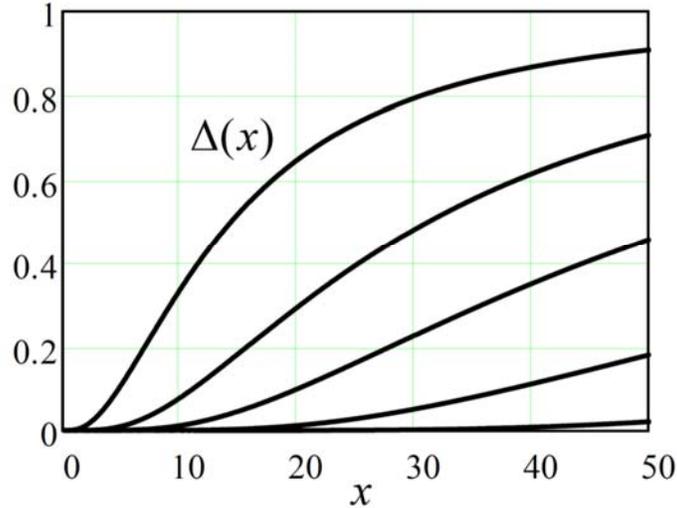

**Fig. 1**: Relative deviation of the representations (18a) from the exact for an arbitrary choice of values of the parameters $a$ and $b$ and for several finite number of terms $N$ in the sum. The relative deviation is defined as $\Delta(x) = \frac{{}_1F_1(a;b;x) - F(x)}{{}_1F_1(a;b;x) + F(x)}$ where $F(x)$ is (18a). We took $a = 3$, $b = 2$, and $N = 3,5,7,10,15$ from top trace to bottom.

The representation (18) is not found in the published literature. It is a viable alternative to another well-known formula that reads (see, for example, page 284 in [4])

$$_1F_1(a;b;z) = 2^{b-1}\Gamma(b)e^{z/2}(2\mu z)^{(1-b)/2}\sum_{n=0}^{\infty} R_n(2\mu z)^{n/2} J_{n+b-1}(\sqrt{2\mu z}), \qquad (19a)$$

where $\mu = b - 2a = y/2i$ and

$$4\mu^2(n+1)R_{n+1} = (n+b-1)R_{n-1} - \frac{1}{2}R_{n-2}, \qquad (19b)$$

for $n = 0,1,2,...$ with $R_0 = 1$, $R_{-1} := 0$, and $R_{-2} := 0$. We should note that it is a highly non-trivial task to use any of the known Bessel function transformations to derive (18) from (19). This is due to the dissimilar nature of the sets $\{P_n\}_{n=0}^{\infty}$ and $\{R_n\}_{n=0}^{\infty}$, which satisfy very different recursion relations. In fact, to achieve that, one has to show that

$$\sum_{n=0}^{\infty} P_n J_{n+\frac{b-1}{2}}(iz/2) = \frac{\Gamma(b)}{\Gamma(\frac{b+1}{2})}(-2i\mu)^{(1-b)/2}\sum_{n=0}^{\infty} R_n(2\mu z)^{n/2} J_{n+b-1}(\sqrt{2\mu z}). \qquad (20)$$

On the other hand, numerically one can easily verify that the representation (18) converges more rapidly than (19). Figure 2 is a pictorial illustration that supports this conclusion. It is a plot of the relative deviation of the representations (18) and (19) from the exact $_1F_1(a;b;z)$.

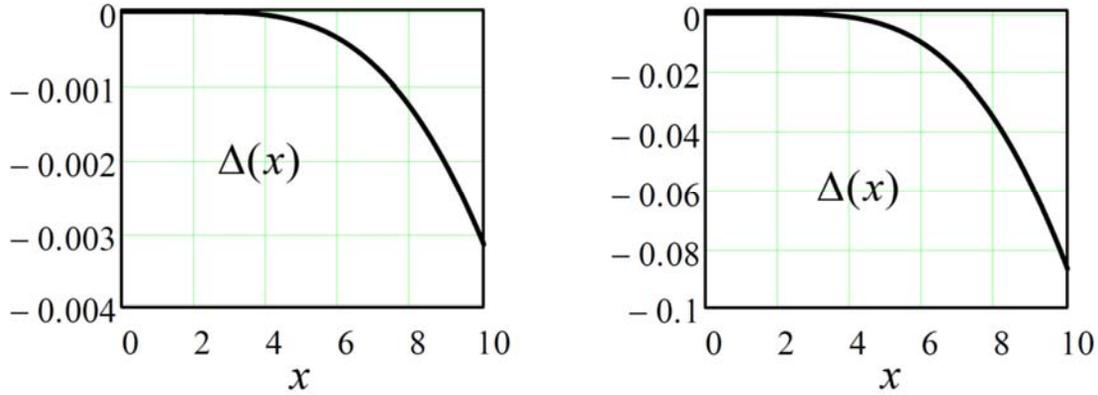

**Fig. 2**: The relative deviation of the representations (18) and (19) from the exact for an arbitrary choice of values of the parameters $a$ and $b$ and for a finite number of terms $N$ in the sum. The relative deviation $\Delta(x)$ is as defined in the caption of Figure 1. We took $a = 2.5$, $b = 3.7$, and $N = 5$.

Finally, Table I is a numerical sample comparing the two representations to the exact values. The Table demonstrates accuracy of the series (18) and its superior convergence over (19), especially asymptotically.

**Table I**: The result of evaluating the representations (18) and (19) as compared to the exact values of $_1F_1(a;b;x)$ to within 12 decimal places of accuracy. We took $a = 2.5$, $b = 3.7$, and $N = 20$. Deviations of digits from the exact are shown in bold.

| x | Exact | Eq. (18) | Eq. (19) |
|---|---|---|---|
| 1.0 | 2.009 719 470 686 | 2.009 719 470 686 | 2.009 719 470 686 |
| 2.0 | 4.205 949 449 938 | 4.205 949 449 938 | 4.205 949 449 938 |
| 3.0 | 9.109 529 330 045 | 9.109 529 330 045 | 9.109 529 330 045 |
| 4.0 | 20.301 955 333 864 | 20.301 955 333 864 | 20.301 955 333 864 |
| 5.0 | 46.326 312 197 243 | 46.326 312 197 243 | 46.326 312 197 24**2** |
| 6.0 | 107.787 310 993 028 | 107.787 310 993 028 | 107.787 310 99**2 981** |
| 7.0 | 254.858 336 524 261 | 254.858 336 524 261 | 254.858 336 52**2 166** |
| 8.0 | 610.734 138 079 992 | 610.734 138 079 99**5** | 610.734 138 0**19 753** |
| 9.0 | 1480.110 669 501 183 | 1480.110 669 501 1**62** | 1480.110 66**8 255 821** |
| 10.0 | 3621.413 368 129 457 | 3621.413 368 129 **395** | 3621.413 3**48 288 315** |

## Acknowledgments

We are grateful to I. A. Assi, A. Jellal, and S. M. Al-Marzoug for the numerical verification of formula (18) using Python and Mathematica codes.## Data Availability Statement:

Data sharing is not applicable to this article as no new data were created or analyzed in this study.